\documentclass[12pt,a4paper,dvips]{article}
\usepackage{epsfig}
\renewcommand\thesection{\arabic{section}.}
\renewcommand\thesubsection{\thesection\arabic{subsection}.}
\renewcommand\thesubsubsection{\thesubsection\arabic{subsubsection}.}
\renewcommand\section[1]{\vspace{\topsep}\vspace{\partopsep}
\refstepcounter{section}
{\par  \noindent\normalsize\bfseries \thesection
\hspace{1em}#1\vspace{\topsep}\par\noindent}}

\newenvironment{refs}
{\vspace{\topsep}\vspace{\partopsep}
{\par \noindent\normalsize\bfseries  References
\vspace{-\topsep}\par\noindent}
\setlength{\parindent}{-5mm}
\begin{list}{}{\topsep 0pt \partopsep 0pt \itemsep 0pt \leftmargin 5mm
\parsep 0pt \itemindent -5mm}}
{\end{list}}

\renewcommand\subsection[1]{
\refstepcounter{subsection}
{\par \protect\vspace{\topsep}\vspace{\partopsep}
 \noindent\normalsize\bfseries \slshape \thesubsection
\hspace{1em}#1\par \noindent}}

\renewcommand\subsubsection[1]{
\refstepcounter{subsubsection}
{\par \protect \vspace{\topsep}\vspace{\partopsep}
\noindent\normalsize \slshape \thesubsubsection
\hspace{1em}#1\par \noindent}}

\newfont{\sansb}{cmssbx10}
\newfont{\sans}{cmss10}

\begin{document}
\begin{center}
{\large \bf Performances of the CAT imaging telescope
and some preliminary results on Mkn 180 and the Crab\vspace{18pt}\\}
{P. Goret$^1$ for the CAT collaboration\vspace{12pt}\\}
{\sl 
$^1$DAPNIA/SAp, CE Saclay, France\\
}
\end{center}
\begin{abstract}
The CAT imaging telesope, which uses the Atmospheric Cherenkov Technique to observe TeV gamma-ray sources, has been operating since September 1996. Located in southern France, it features a fine grained camera consisting of 546 PMT with a pixel size of 0.12$^\circ$. Strong gamma-ray signals from Mkn 501 above $\sim$200 GeV were detected during observations made in April 1997. The high signal/noise ratio for these data enables direct characterization of the gamma-ray images as recorded in the camera. These results are used to validate the Monte-Carlo simulations of the response of the telescope to gamma-ray showers and to improve its performances. Observational results on Mkn 180 and the Crab nebula spectrum are presented.
\end{abstract}
\setlength{\parindent}{1cm}
\section{Introduction}
The CAT imaging telescope has been described elsewhere (Rivoal, 1997). Briefly, it consists of a 18 m$^2$ mirror and a 546 PMT camera with a pixel size of 0.12$^\circ$. The trigger is generated out of the 288 inner PMT's, which are further divided into 9 overlapping sectors of 48 PMT's each. 
The trigger condition is that at least 4 PMT's in any sector fire above a threshold of 3 photoelectrons.
The PMT signals, gated by a 12 ns wide analog switch, are recorded by a 15 bits charge ADC. The trigger rate is of the order of $\sim$15 Hz at zenith. Extensive Monte-Carlo simulations of gamma-ray Cherenkov showers were performed in order to determine the image parameters (width, length, orientation) and effective collecting area as a function of energy and zenith angle. These results are used to i) optimize the image cuts for gamma/hadron discrimination and ii) estimate the absolute fluxes and energy spectra from gamma-ray sources. With the absence of a calibration gamma-ray beam, one must rely on the best possible knowledge concerning shower development, atmospheric properties, optical parameters and electronic signal processing. Only recently, the intense flaring activity of Mkn 421 and Mkn 501 enabled the detection of several thousands of gamma-rays with a high signal/noise ratio. These data can be used to check measured gamma-ray image parameters against MC simulations. In this paper, the distributions of width and length for gamma-rays in different energy ranges, as measured during strong flares from Mkn 501, are presented and compared with the results of simulations. Energy-dependant image cuts are derived which optimize the gamma/hadron separation. These results are used to search for a gamma-ray signal from Mkn 180 with improved sensitivity. Finally the differential energy spectrum of the Crab nebula in the energy range 0.4-5.0 TeV is presented.
\eject

\section{Image parameters : real data vs. MC}
During observations in the night of April 16 1997, an intense flare from the direction of Mkn 501 was detected by the CAT telescope. Showing a high signal/noise ratio, these data were used to study the gamma-ray image parameters as recorded by the CAT telescope at zenith angles $\le$30$^\circ$. The moment-based parameters {\it width} and {\it length} were calculated for each event in both the ON and OFF data.  Events were selected for analysis according to the following cuts: {\it alpha}$\le$10$^\circ$ and 0.5$^\circ \le${\it dist}$\le$1.3$^\circ$. The results were sorted according to a parameter related to the light intensity, viz. {\it logsize}=log({\it size}) where {\it size} is the total number of photoelectrons recorded in the image. Subtracting the distributions of width and length in the ON data from those in the OFF data, these distributions pertaining to gamma-rays were obtained. The observed width and length distributions for gamma-rays are shown in Figs. 1 and 2 respectively for different bands in {\it logsize}, together with the same distributions for hadronic background events (OFF data). It is clearly seen that the gamma/hadron separation increases dramatically with increasing {\it logsize}. In the lowest {\it logsize} band [1.5-2.0], one can see rather poor discrimination between gamma-rays and hadrons. The optimum sensitivity, trading between $\gamma$/h discrimination and statistics, is obtained by selecting events with {\it logsize}$\ge$1.9 corresponding to an energy threshold of $\sim$400 GeV. The resulting sensitivity for the Crab is 4.6$\sigma$t$^{1/2} _{hours}$ with a detection rate of 1.5 $\gamma$/minute. 

Also shown in figs. 1 and 2 are the results from MC simulations. The agreement between observed and simulated distributions is seen to be quite satisfactory, thus strenghtening the confidence in MC simulations which are further used to estimate energy spectra and absolute fluxes. The optimized values for the cuts on {\it width} and {\it length} are given in Table 1. According to MC, these cuts remove only $\sim$20\% of gamma-rays.

\begin{table}[hbt]
\caption{The optimized cuts on width and length in different {\it logsize} bands (in degrees)}\label{Topic}
\begin{center}
\begin{tabular}{|c|c|c|c|c|}
\hline
{\it logsize} & wmin & wmax & lmin & lmax \\
\hline
1.5-2.0 & 0.010 & 0.076 & 0.060 & 0.290 \\
2.0-2.5 & 0.018 & 0.100 & 0.118 & 0.329 \\
2.5-3.0 & 0.029 & 0.123 & 0.138 & 0.353 \\
3.0-5.0 & 0.059 & 0.176 & 0.176 & 0.412 \\
\hline
\end{tabular}
\end{center}
\end{table}

\section{Observations of Mkn 180}
The optimized cuts for the moment-based analysis, as derived in the preceding section, were used to search for a gamma-ray signal from Mkn 180. The database consists of 15 ON/OFF pairs with a total of 5.4 hours each at zenith angle $\le$30$^\circ$ taken in March and April 1997. After applying the optimizedcuts ({\it alpha}$\le$10$^\circ$, {\it logsize}$\ge$1.9, width and length as given in Table 1), the remaining number of events are N$_{ON}$=1131 and N$_{OFF}$=1142 for the ON and OFF data respectively. The resulting 3$\sigma$ upper limit for the gamma-ray flux from Mkn 180 is 0.44 $\gamma$/minute or, equivalently, 0.29 Crab unit.

\section{The Crab nebula differential energy spectrum}
The Crab nebula differential energy spectrum was derived using 8.5 hours on-source observations in the ON/OFF mode made during the 1996/97 winter season. The energy assigned to each event was estimated according to the method elaborated by Le Bohec (1996). The method relies on fitting each recorded Cherenkov image to a template of images generated at different gamma-ray energies and impact parameters to the telescope. A chi-square cut to the fitted events selects gamma-like events. According to MC simulations, the energy resolution is of the order of 20-25\%. MC simulations are also used to estimate the effective collecting area and spill-over for each energy bin. The resulting differential energy spectrum is $\Phi_{CRAB}$=2.46$\pm$0.22$\times$10$^{-7}$E$^{-2.55\pm 0.09}$m$^{-2}$s$^{-1}$TeV$^{-1}$ with the errors quoted being statistical only. Correction for an instrumental deadtime of 7\% was taken into account. The Crab differential energy spectrum is shown in Fig. 3.

\section{Conclusions}
A study of intense gamma-ray flares from Mkn 501, as observed by the CAT imaging telescope, has led to optimize the moment-based analysis cuts. These results were also used to validate the MC simulation of the experiment. A 3$\sigma$ upper limit to the gamma-ray flux from Mkn 180 was derived. Finally, the differential energy spectrum of the Crab nebula in the 0.4-5.0 TeV range was presented. More investigations are being made to further improve the data analysisand, particularly, to study the systematic errors.. Recently, an additional guard ring of 54 PMT's was implemented in the CAT camera which must lead to better energy resolution.

\begin{refs}
\item Rivoal M. 1997, Proc. 25$^{th}$ ICRC Conf., OG 10.3 12, Vol. 5, p. 89.
\item Le Bohec S. 1996, PhD. thesis, Paris. 
\end{refs}



\begin{figure}[hb]
\centerline{\epsfig{file=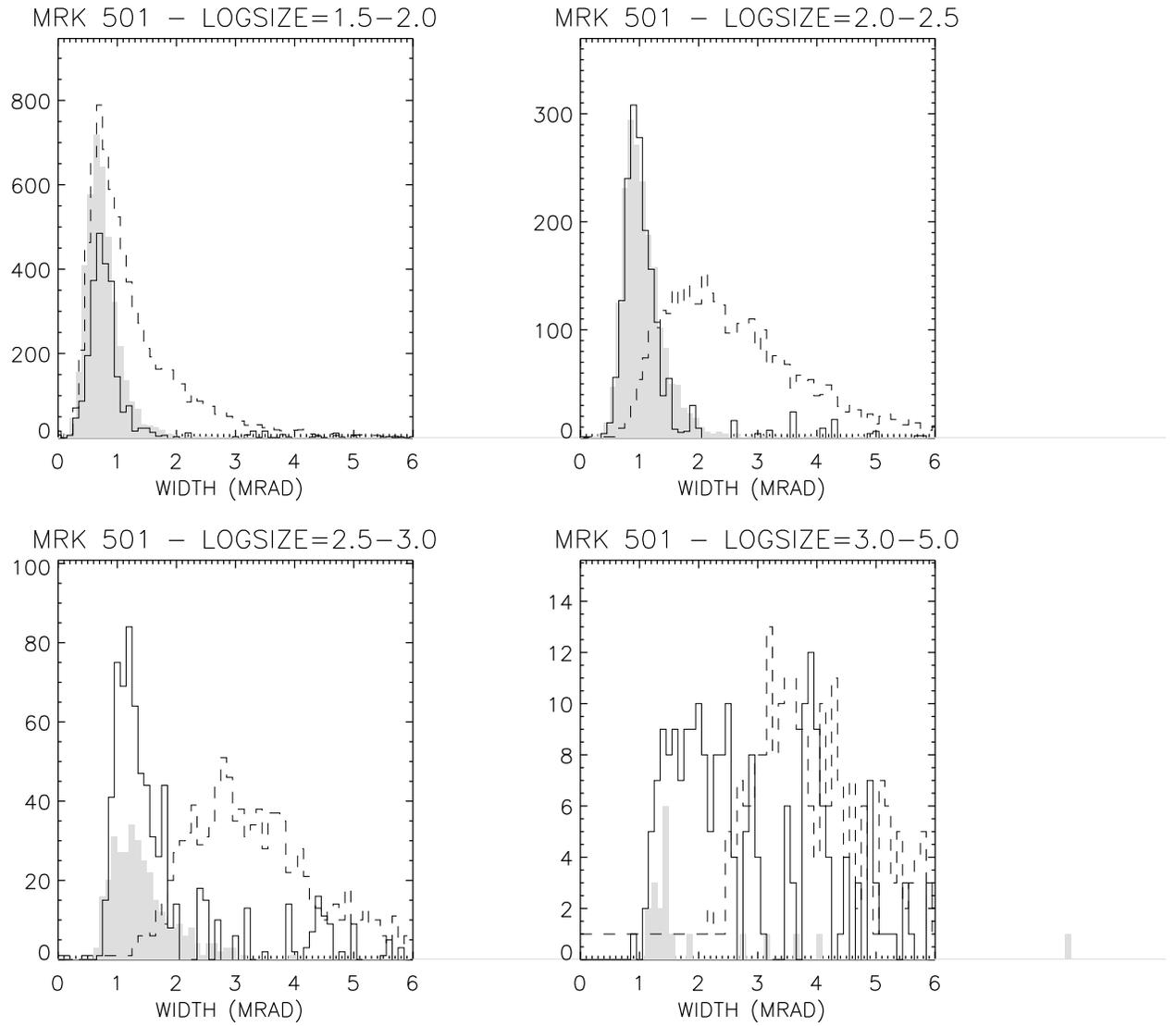,height=15.0cm}}
\caption{Distributions of the {\rm width} parameter for $\gamma$-rays from Mkn 501 (solid line), for hadronic background (dashed line) and MC $\gamma$-ray simulations (grey histogram) in 4 different {\rm logsize} bands (see text).}
\end{figure} 

\begin{figure}[hb]
\centerline{\epsfig{file=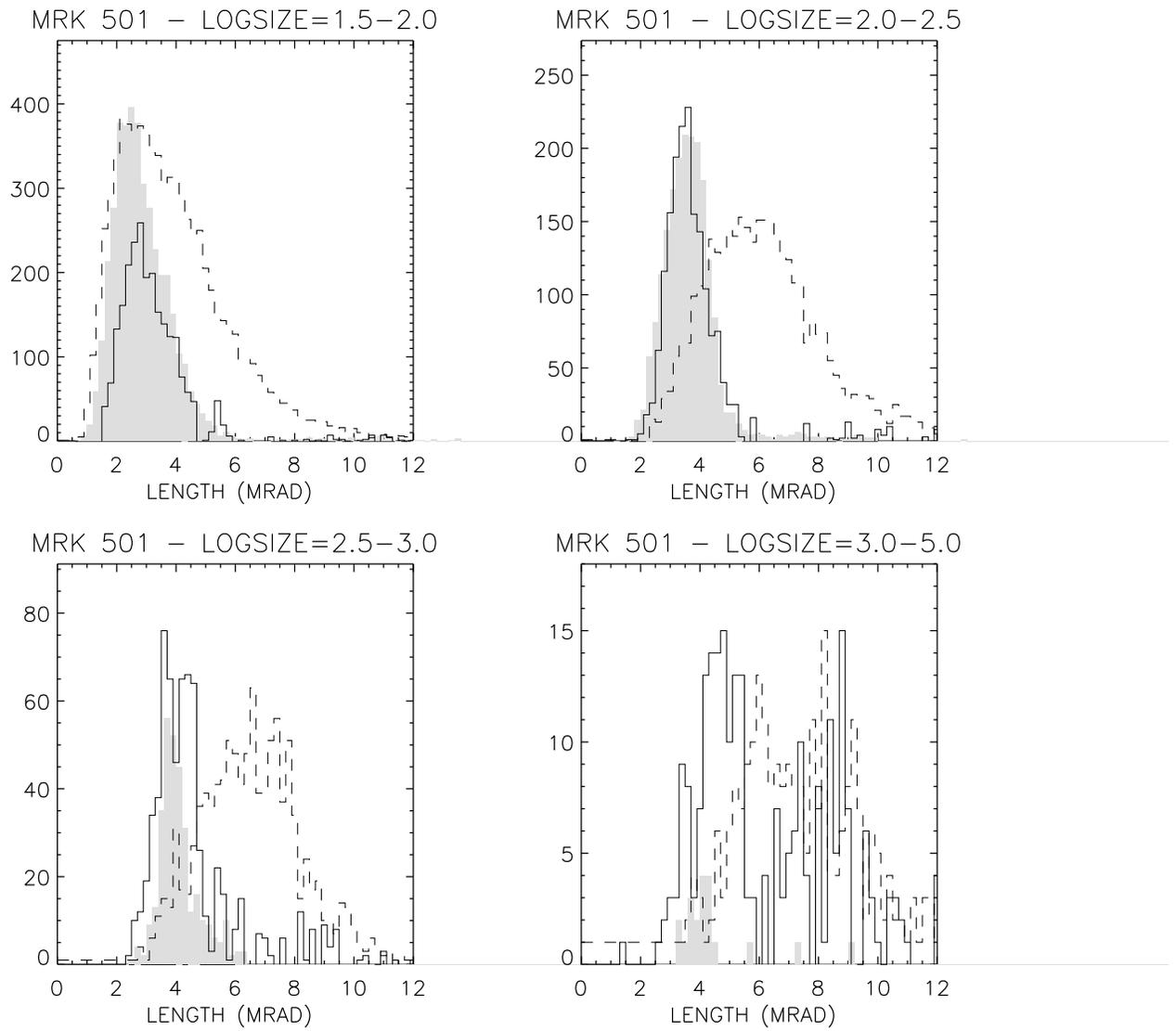,height=15.0cm}}
\caption{Same as Fig. 1 for the {\rm length} parameter.}
\end{figure} 

\begin{figure}[hb]
\centerline{\epsfig{file=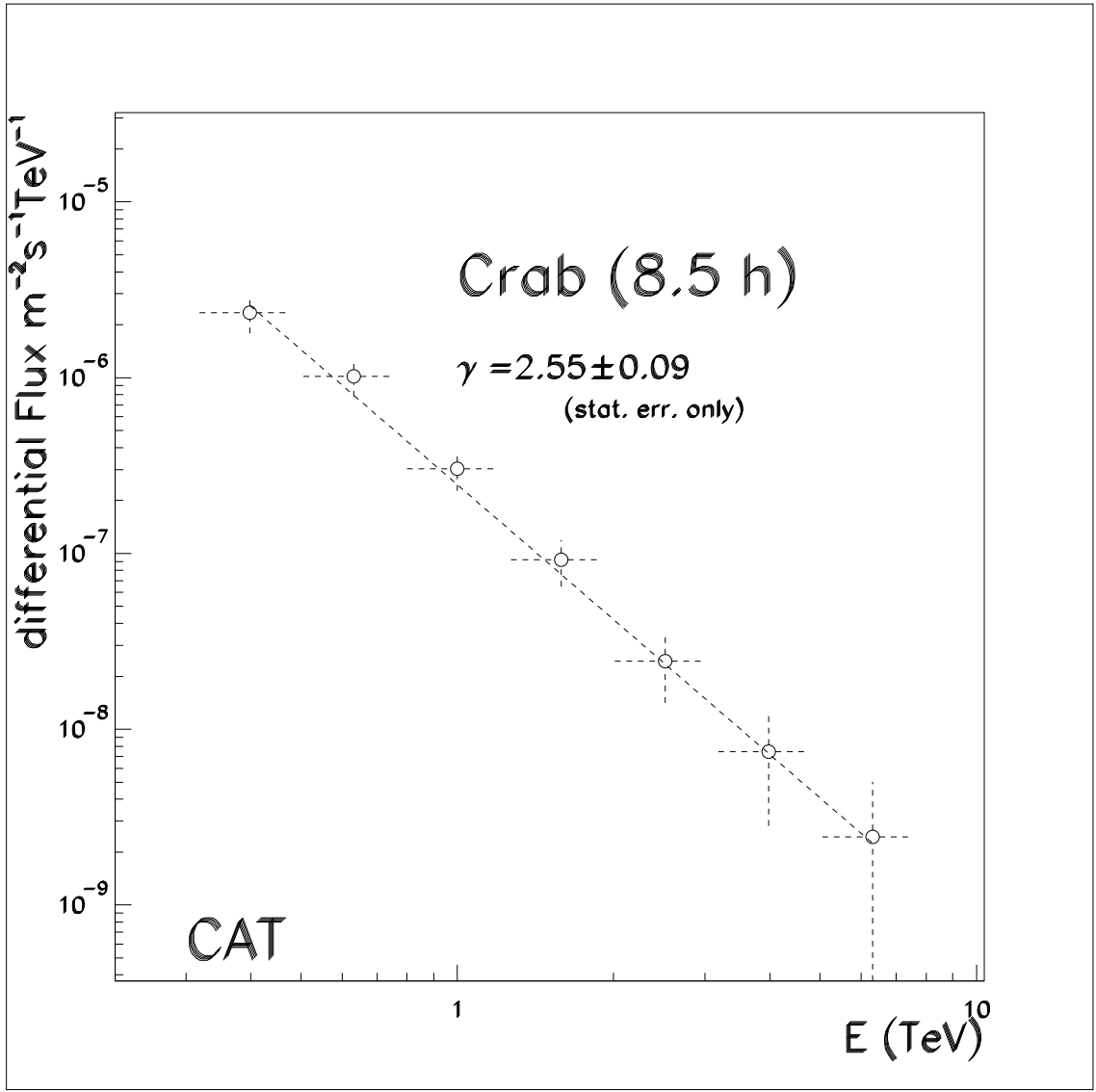,height=15.0cm}}
\caption{Differential spectrum of the Crab nebula.}
\end{figure}

\end{document}